# Origin of the excess specific heat in metallic glass forming liquids

Hai Bo Ke, Ping Wen*, Wei Hua Wang

Institute of Physics, Chinese Academy of Sciences, Beijing 100190, P. R. China

ABSTRACT

We report excess specific heat in a series of metallic glass forming liquids. It is found that the excess specific heat relative to glass at glass transition temperature $T_g$ is constant and close to $\frac{3}{2}R$, where $R$ is gas constant. In the typical $Pd_{40}Ni_{10}Cu_{30}P_{20}$ metallic glass forming liquid, the excess specific heat is independent of temperature. A quantitative description of the excess specific heat is built up. The atomic translational diffusion is the origin of the excess specific heat. The results provide a fundamental understanding to the glass transition in metallic glasses.

Contact author, email: pwen@aphy.iphy.ac.cn

Excess specific heat, $\Delta C_p$, of glass forming liquids relative to their rigid glass or crystals at ambient pressure is of particular importance to understand the nature of the glass transition [1-4]. It is the basis for the Kauzmann paradox [1] that indicates some type of phase transition must occur between the liquid and the glass. But whether the glass transition is a phase transition or not is still controversial [5-9]. The glass transition is universal. Without crystallization most of liquids (including molecular, ionic and metallic types) can be undercooled and solidified gradually into glass across the glass transition temperature. In molecular GFLs, the $\Delta C_P$ has been described popularly in terms of "bead" [4]. "Bead" is atomic group in a molecule. Its complicated rotations can change the configurational state of the system. In most of molecular GFLs, $\Delta C_P$ is an approximately constant around 11.2 Jmole$^{-1}$bead$^{-1}$ [5]. But an ambiguity is found when comparing rigidly with different molecules where the same number of atoms is bound together in different ways [10]. Conventionally, the $\Delta C_p$ of GFLs is ascribed to configurational rearrangements in glass forming liquids (GFLs) [3]. It corresponds to the long-standing assumption for the motions in GFLs, in which vibrations on the short time scale are separated from configurational rearrangement on the long time scale [11]. The configrational rearrangement in GFLs has been referred as the thermal fluctuation of cooperative region but individual atomic and molecular motion in GFLs [5, 12]. The conventional wisdom provides a qualitative description to the empirical pattern of the $\Delta C_p$ in GFLs. The $\Delta C_p$ increases gradually as temperature approaches the glass transition temperature $T_g$ from high temperature [10, 13], and fragile GFLs have a larger $\Delta C_p$ at $T_g$ than strong GFLs have [14]. Recently this pattern has been challenged in a broaden spectrum of GFLs [15]. Up to now, it is still lack of an agreement on the origin of the $\Delta C_P$ in GFLs.

Metallic GFLs are paradigms of the dense random packing of spheres, hence of great fundamental interest in studying of the nature of the glass transition [16]. Until recently to study thermodynamic and dynamic properties in a serial of metallic supercooled liquids is not possible since the conventional metallic glasses have a strong tendency to crystallize when heated through the glass transition. Current development of bulk metallic glass forming systems [17] has offered a good opportunity to study metallic GFLs. The dynamic characteristics [17, 18] and the diffusion mechanism [19] of metallic GFLs have been studied in metallic GFLs. Their structural relaxation has non-Arrhenius and non-exponential behavior that similar to those in most GFLs [9]. Atomic collective motion is a dominant process in metallic GFLs. An obvious $\Delta C_P$ has been observed in several metallic GFLs [20]. But a clear pattern of the $\Delta C_P$ for metallic GFLs is still missing. The origin of the $\Delta C_P$ of metallic GFLs related to the nature of the glass transition is still beyond our understanding.

In this letter, we report the $\Delta C_P$ in a series of metallic glass forming liquids. The measurements of the $\Delta C_P$ and the specific heat in these metallic GFLs was carried out in a Perkin-Elmer DSC7 and Mettler Toledo DSC 1 thermal analyzer at the heating rate or cooling rate of 20 K/min with an error less than 10%. The $\Delta C_P$ relative to glass at glass transition temperature $T_g$ is found to be constant and close to $\frac{3}{2}R$ ($R$ is gas constant). The $\Delta C_P$ in typical $Pd_{40}Ni_{10}Cu_{30}P_{20}$ metallic glass forming liquid is independent of temperature. Based on the correlation between the $\Delta C_P$ and the atomic translational diffusive degrees in these GFLs, a quantitative description to the $\Delta C_P$ in metallic GFLs is provided.

Figure 1 plots the $\Delta C_p$ reduced by $\frac{1}{2}R$ for 45 kinds of metallic GFLs. The $\Delta C_P$ is the specific heat difference between the supercooled liquid and glass at $T_g$ [19]. The names of the metallic glass formers are listed in Table I. One can find that the value of $\frac{2\Delta C_P}{R}$ is distributed into a narrow region from 3 to 3.3, while their $T_g$ has a wide distribution from 300 to 700 K. Within the experimental error the $\Delta C_P$ of metallic GFLs seems to be a constant and close to $\frac{3}{2}R$. The constant $\Delta C_P$ at $T_g$ for metallic GFLs is not consistent with to the empirical pattern of the $\Delta C_P$ in GFLs [3, 14, 15]. It is due to that the exact dynamics for those metallic GFLs in Table I are different. The values of their fragility are distributed from 30 to 60 [21]. The idea of configurational rearrangement can not explain the constant $\Delta C_P$ at $T_g$ in these metallic GFLs. In term of "bead" [4] the $\Delta C_P$ of the metallic GFLs should be $\frac{3}{2}R$ per bead (12.5 $JK^{-1}bead^{-1}$) because metallic bond is dominative in the metallic GFLs and each atom in the metallic system can be considered directly as a "bead". Some values of the $\Delta C_p$ in $JK^{-1}bead^{-1}$ are summarized as follows: selenium 13.5 [22], methanol 11 [23], *o*-terphenyl 37 [24] and isopentance 17 [25]. It does not seem to yield a relatively constant of $\Delta C_P$ in GFLs. No theory has been found to describe the ambiguity of the $\Delta C_P$ in term of "bead" yet.

It is usual, in theoretical considerations, to discuss specific heat per mole atom at constant volume. Fortunately, the excess specific heat, $\Delta C_V$, at constant volume in metallic GFLs can be considered to be close to their $\Delta C_P$ with ignoring the expansion and the consequent changes in elastic properties. This neglect involves only small errors (probably of a few per cent), and is similar as that the specific heat at constant pressure is close to that at constant volume for a metallic glass [26]. In order to clarify the origin of $\Delta C_V$ or $\Delta C_P$ at $T_g$, one must understand the difference of the intrinsic motions in metallic GFLs and glasses. In nature the specific heat of a condensed matter must be related directly to its intrinsic motions. It is known that structural relaxation is accessible in metallic GFLs but their glasses on the experimental time scale [18]. For metallic GFLs the existence of atomic translational diffusion, i.e., atomic jumps between potential energy minima on the long time scale, is necessary to the structural relaxation. This means atomic translational diffusions in metallic GFLs are accessible during normal heating or cooling on the experimental scale beyond the vibrations of atoms in the potential energy minima. Moreover, the dominative metallic bond makes it reasonable that atomic diffusions in the metallic GFLs are equivalent for the internal energy of the system.

The internal energy related to atomic translational diffusion in metallic GFLs has a form: $E\left(\vec{p}\right) = c\,\vec{p}^{\,2}$, where $c$ is a constant coefficient and $\vec{p}$ is the momentum vector for an atomic translational diffusion. In space, the above equation can be written as $E\left(\vec{p}\right) = c\,\vec{p}^{\,2} = \sum_{i=1}^{3} c_i p_i^2$, where $c_i$ is constant related to $p_i$. The $p_i$ is the momentum variable, and equivalent to $p_x$, $p_y$, and $p_z$ with Cartesian coordinates *x, y*, and *z*. Here, classical mechanics can be used to analyze the average internal energy $\overline{E}$ per atom in metallic GFLs. Each value of $p_i$ corresponds to a separate, independent state. The states are discretely spaced, separated by small intervals $\Delta p$ because the atoms in metallic GFLs have random positions in space. Considering that $\Delta p$ is much less than *kT* (*k*, Boltsmann constant, *T*, temperature of system), one can write the partition function related to the atomic translational diffusion as following form: $Z = \sum_p e^{-\frac{E(p)}{kT}} = \sum_{p_x} e^{-\beta c_1 p_x^2} \sum_{p_y} e^{-\beta c_2 p_y^2} \sum_{p_z} e^{-\beta c_3 p_z^2}$, where $\beta$ is *kT*. To evaluate the product, the $\sum_p e^{-\beta c p^2}$ in above equation is transformed into the form: $\frac{1}{\Delta p}\sum_p e^{-\beta c p^2}\Delta p = \frac{1}{\Delta p}\int_{-\infty}^{+\infty} e^{-\beta c p^2} dp = \frac{1}{\Delta p}\frac{1}{\sqrt{\beta c}}\int_{-\infty}^{+\infty} e^{-u^2} du$. It has turned out mathematically that the integral of $e^{-u^2}$ from -∞ to +∞ is exactly equal to $\sqrt{\pi}$. So the final result for the partition function is $Z = \left(\frac{1}{\Delta p}\sqrt{\frac{\pi}{\beta c}}\right)^3 = C\beta^{-3/2}$, where *C* is just an abbreviation for $\left(\frac{\sqrt{\pi/c}}{\Delta p}\right)^3$. It is convenient to calculate the $\overline{E}$, using the formula: $\overline{E} = -\frac{1}{Z}\frac{\partial Z}{\partial \beta}$. The value of $\overline{E}$ per atom in metallic GFLs is $\frac{3}{2}kT$. So the $C_V$ per mole atoms contributed from the translational diffusion in metallic GFLs can be derived from the equation ($C_V = N_0 \frac{d\overline{E}}{dT}$, $N_0$ is Avogadro's number). It is found that the $C_V$ ($\frac{3}{2}R$) contributed from the translational diffusion is equal exactly to the $\Delta C_P$ of metallic GFLs at $T_g$.

It is not coincident that the $C_V$ arisen from the atomic translational diffusion is equal exactly to the $\Delta C_P$ of metallic GFLs at $T_g$. With the conventional idea [3, 8], in viscous metallic liquids, the vibration is separated from the structural rearrangement because the rate of the structural relation is lower by many orders of the magnitude than a characteristic vibration frequency. It is plausible intuitively that the intrinsic atomic motions contributing to the specific heat in metallic GFLs with the structure of the dense random packing of spheres [16] involve only two types: atomic vibrations and translational diffusions. The late is the one contributing to the $\Delta C_P$ of metallic GFLs since the former can exist in both of glass and supercooled liquid. That is, the contribution to the specific heat in metallic GFLs can be separated into vibrational and translational diffusive parts. The $\Delta C_P$ in metallic GFLs originates from the translational diffusions that obey the equipartition theorem. Its value is close to $\frac{f}{2}R$, where $R$ is gas constant. $f$ is the total number of translational degrees of freedom. For metallic GFLs the value of $f$ is 3 and equal to the number of independent coordinates for atomic translational diffusions in space.

The correlation between the $\Delta C_P$ and translational degrees indicates that the $\Delta C_P$ in metallic GFLs is independent of temperature. As a characteristic of translational diffusive motion, the translational degrees for a metallic GFL is not dependent of the temperature. The value of the specific heat related to atomic translational diffusion is just $\frac{3}{2}R$. At the same time, one can evaluate the vibrational contribution to the specific heat in metallic GFLs. It is due to the fact that Debye temperature of metallic glasses is always well less than their $T_g$ [27], indicating all atomic vibrations have already been activated in metallic glasses at temperature above $T_g$. Then the contribution of atomic vibration to the specific heat in metallic GFLs is similar to that in case of an Einstein solid in the high temperature limit [28]. The specific heat arisen from atomic vibrations is independent of temperature, and has the value of 3$R$. The sum of the two parts in metallic GFLs is 4.5$R$ (37.4 $JK^{-1}mol^{-1}$), and be temperature independent. This is consistent with the results in Figure 2. The specific heat of $Pd_{40}Ni_{10}Cu_{30}P_{20}$ metallic GFL with good glass forming ability [29] is of temperature independence, and around 40.5 $JK^{-1}mol^{-1}$ within the experimental error.

The origin of the excess specific heat in these typical metallic GFLs provide a fundamental understanding to the glass transition. In metallic GFLs each atom can diffuse translationally and freely within the experimental time. Upon cooling, the glass transition in metallic glass formers accompanied by a disappearance of the excess specific heat is directly related to the frozen of the translational diffusions for almost all of atoms in metallic GFLs. The glass transition is a pure kinetic process. Since no phase transformation can be related to the frozen of atomic translational diffusions in metallic GFLs, the Kauzmann paradox [1], implying that the glass transition is an underlying phase transformation, is not still a problem. The slowing down of the metallic GFLs, which represents the increase in shear viscosity by several orders in magnitude [17], can be related to the increase in the average activation energy for the translational diffusive motions. Similar as the description in the elastic model of the glass transition [8], the temperature dependence of the average structural relaxation time is non-Arrehenius. Due to the positions of each atom in metallic GFLs are random in space, the activation energy for translational diffusions is a wide distribution. The broaden distribution of the activation energy has been used widely to describe the nonexponential structural relaxation in GFLs [29].

In summary, differential scanning calorimeter is used to measure the excess specific heat in a serial of metallic glass forming liquids. The excess specific heat at $T_g$ is found to be equal to $\frac{3}{2}R$, where $R$ is gas constant. $Pd_{40}Ni_{10}Cu_{30}P_{20}$ melt, a typical metallic GFL, has a temperature-independent specific heat. A correlation between the excess specific heat and the number of translational degree in metallic GFLs is built up. These results show the glass transition in metallic glasses is a pure kinetic process.

The authors thank C. Austen Angell (Arizona State University) and Takeshi Egami (University of Tennessee) for fruitful discussions. Financial supports from the Science Foundation of China under Grant 51071170, 50731008, and 50921091 and from MOST 973 of China under 2007CB613904 and 2010CB731603 are gratefully acknowledged.

**References:**

1. W. Kauzmann, Chem. Rev. **43**, 219 (1948).
2. M. Goldstein, J. Chem. Phys. **64**, 4767 (1965).
3. J. Jackle, Rep. Prog. Phys. **49**, 171 (1986).
4. B. Wunderlich, J. Chem. Phys. **64**, 1052 (1960).
5. G. Adam, J. H. Gibbs, J. Chem. Phys. **43**, 139 (1965).
6. M. H. Cohen, G. S. Grest, Phys. Rev. B **20**, 1077 (1979).
7. G. H. Fredrickson, H. C. Andersen, Phys. Rev. Lett. **53**, 1244 (1984).
8. J. C. Dyre, N. B. Olsen, T. Christensen, Phys. Rev. B **53**, 2171 (1996).
9. E. Donth. *The glass transition: Relaxation dynamics in liquids and disordered material* (Spring-Verlag Berlin Heidelberg, 2001).
10. C. Alba-Simionesco, J. Fan, C. A. Angell, J. Chem. Phys. **110**, 5262 (1999).
11. M. Goldstein, Faraday Symp. Chem. Soc. **6**, 7 (1969); F. H. Stillinger, T. A. Weber, Phys. Rev. A **28**, 2408 (1983).
12. J. H. Gibbs, E. A. DiMarzio, J. Chem. Phys. **28**, 373 (1958).
13. H. Tanaka, Phys. Rev. Lett. **90**, 55701 (2003).
14. C. A. Angell, in *relaxations in complex systems*, ed. L. Ngai, G.B. Wright (U. S. GPO, Washington, D. C., 1985), p.3; U. Mohanty, *Advances in Chemical Physics*, ed. I. Prigogine, S. A. Rice (Wiley, New York) **89**, 89 (1995); X. Xia, P. G. Wolynes, Proc. Natl. Acad. Sci. U.S.A. **97**, 2990 (2000).
15. D. Huang, G. B. McKenna, J. Chem. Phys. **114**, 5621 (2001).
16. J. D. Bernal, Nature **185**, 68 (1960); R. W. Cahn, in *Materials Science and Technology*, (VCH, Weinheim), p. 403 (1991).
17. A. L. Greer, Science **267**, 1947 (1995); W. L. Johnson, MRS Bull. **24**, 42 (1999); W.H. Wang, C. Dong, C.H. Shek, Mater. Sci. Eng. **R44**, 45 (2004).
18. P. Wen, D. Q. Zhao, M. X. Pan, W. H. Wang, Y. P. Huang, M. L. Guo, Appl. Phys. Lett. **84**, 2790 (2004).
19. X. P. Tang, U. Geyer, R. Busch, W. L. Johnson, Y. Wu, Nature **402**, 160 (1990).
20. H. B. Ke, P. Wen, D. Q. Zhao, W. H. Wang, Appl. Phys. Lett. **96**, 251902 (2010).
21. Z. F. Zhao, P. Wen, C. H. Shek, W. H. Wang, Phys. Rev. B **75**, 174201 (2007).
22. C. J. B. Clew, K. Lonsdale, Proc. Roy. Soc. (London) **A161**, 493 (1937).
23. I. L. Karle, L. O. Brockway, J. Am. Chem. Soc. **66**, 1974 (1944).
24. S. S. Chang, A. B. Bestul, J. Chem. Phys. **56**, 503 (1972).
25. R. J. Greet, D. Turnbull, J. Chem. Phys. **47**, 2185 (1967).
26. P. Wen, G. P. Johari, R. J. Wang, W. H. Wang, Phys. Rev. B **73**, 224203 (2006).
27. W. H. Wang, P. Wen, D. Q. Zhao, M. X. Pan, R. J. Wang, J. Mater. Res. **18**, 2747 (2003).
28. D. V. Schroeder, *An introduction to thermal physics* (Addison Wesley Longman, 2000).
29. H. Sillescu, J. Non-Cryst. Solids **243,** 81 (1999).

## Captions

Figure 1. (a) Plots of the glass transition temperature $T_g$ *versus* $N$ (the number of the metallic GFLs in Table I), (b) The reduced excess specific heat $2\Delta C_P / R$ ($R$, gas constant) *versus* $N$.

Figure 2. The specific heat for metallic $Pd_{40}Ni_{10}Cu_{30}P_{20}$ glass forming system. Red line is determined during the cooling from the temperature above melting at 20 K/min; black one is measured during the heating at 20 K/min after the above cooling; blue one is measured by the heating of 20 K/min for crystals.

Table I. The excess specific heat, $\Delta C_P$, at $T_g$ for 45 kinds of metallic glass formers. $T_g$ is the onset glass transition temperature measured during the heating at 20 K/min. $R$ is gas constant.

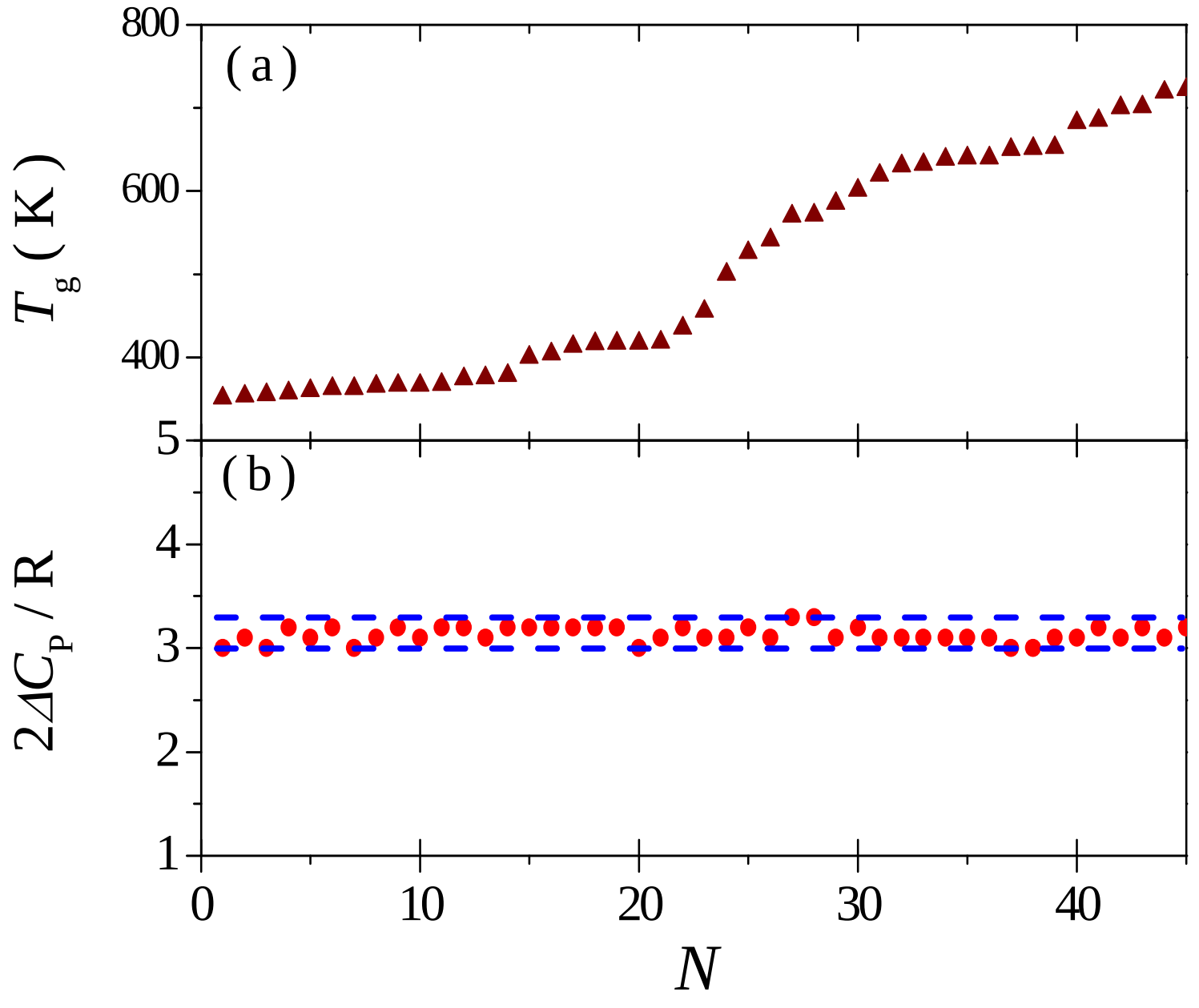

Figure 1. H. B. Ke *et. al*

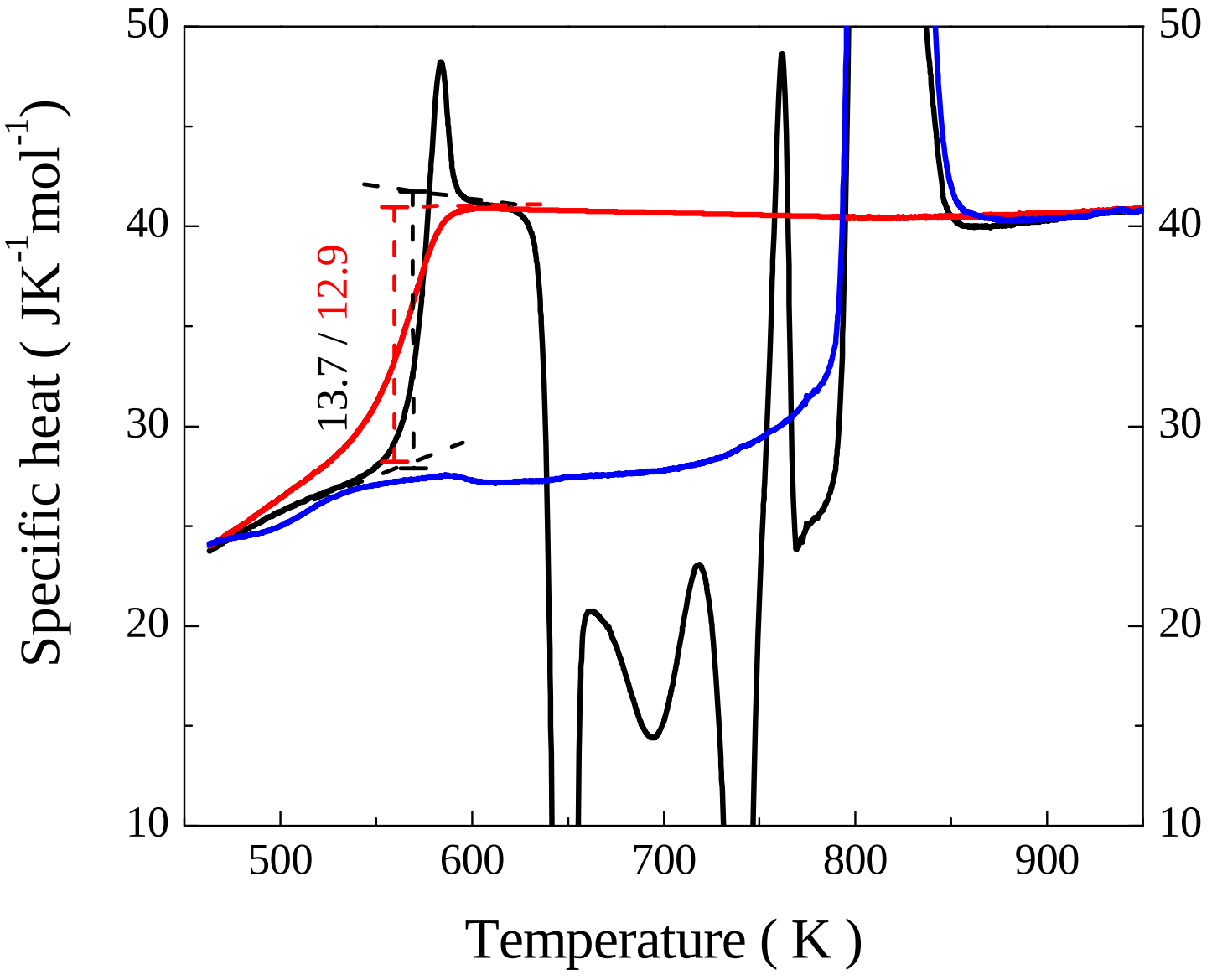

Figure 2. H. B. Ke *et. al*

Table I. H. B. Ke *et. al*

| $N$ | Name | $T_g$ (K) | $\Delta C_P$ (JK$^{-1}$mol$^{-1}$) | $2\Delta C_P/R$ | $N$ | Name | $T_g$ (K) | $\Delta C_P$ (JK$^{-1}$mol$^{-1}$) | $2\Delta C_P/R$ |
|---|---|---|---|---|---|---|---|---|---|
| 1 | $Ce_{68}Al_{10}Cu_{20}Co_2$ | 351 | 12.6 | 3.0 | 24 | $Pr_{55}Al_{25}Co_{20}$ | 500 | 12.7 | 3.1 |
| 2 | $(La_{0.1}Ce_{0.9})_{68}Al_{10}Cu_{20}Co_2$ | 353 | 12.7 | 3.1 | 25 | $La_{55}Al_{25}Co_{20}$ | 526 | 13.4 | 3.2 |
| 3 | $(La_{0.2}Ce_{0.8})_{68}Al_{10}Cu_{20}Co_2$ | 355 | 12.6 | 3.0 | 26 | $Nd_{55}Al_{25}Co_{20}$ | 541 | 12.8 | 3.1 |
| 4 | $(La_{0.3}Ce_{0.7})_{68}Al_{10}Cu_{20}Co_2$ | 357 | 13.5 | 3.2 | 27 | $Pd_{40}Ni_{40}P_{20}$ | 570 | 13.7 | 3.3 |
| 5 | $(La_{0.4}Ce_{0.6})_{68}Al_{10}Cu_{20}Co_2$ | 360 | 12.9 | 3.1 | 28 | $Pd_{40}Cu_{30}Ni_{10}P_{20}$ | 571 | 13.7 | 3.3 |
| 6 | $(La_{0.5}Ce_{0.5})_{68}Al_{10}Cu_{20}Co_2$ | 362 | 13.4 | 3.2 | 29 | $Sm_{40}Y_{15}Al_{25}Co_{20}$ | 585 | 12.7 | 3.1 |
| 7 | $(La_{0.6}Ce_{0.4})_{68}Al_{10}Cu_{20}Co_2$ | 362 | 12.5 | 3.0 | 30 | $Gd_{55}Al_{25}Co_{20}$ | 601 | 13.4 | 3.2 |
| 8 | $Au_{60}Cu_{15.5}Ag_{7.5}Si_{17}$ | 365 | 12.9 | 3.1 | 31 | $Tb_{55}Al_{25}Co_{20}$ | 619 | 12.9 | 3.1 |
| 9 | $(La_{0.8}Ce_{0.2})_{68}Al_{10}Cu_{20}Co_2$ | 366 | 13.1 | 3.2 | 32 | $Zr_{58.5}Cu_{15.8}Ni_{12.5}Al_{10.3}Nd_{7.8}$ | 630 | 12.7 | 3.1 |
| 10 | $(La_{0.7}Ce_{0.3})_{68}Al_{10}Cu_{20}Co_2$ | 366 | 12.9 | 3.1 | 33 | $Dy_{55}Al_{25}Co_{20}$ | 632 | 12.8 | 3.1 |
| 11 | $(La_{0.9}Ce_{0.1})_{68}Al_{10}Cu_{20}Co_2$ | 367 | 13.4 | 3.2 | 34 | $Y_{55}Al_{25}Co_{20}$ | 638 | 12.7 | 3.1 |
| 12 | $La_{68}Al_{10}Cu_{20}Co_2$ | 374 | 13.4 | 3.2 | 35 | $Zr_{46.75}Ti_{8.25}Cu_{7.5}Ni_{10}Be_{27.5}$ | 640 | 12.9 | 3.1 |
| 13 | $Ca_{65}Mg_{15}Zn_{20}$ | 375 | 13 | 3.1 | 36 | $Pd_{77.5}Cu_6Si1_{6.5}$ | 640 | 12.7 | 3.1 |
| 14 | $Ce_{62}Al_{10}Cu_{20}Co_3Ni_5$ | 378 | 13.4 | 3.2 | 37 | $Zr_{65}Al_{7.5}Ni_{10}Cu_{17.5}$ | 650 | 12.5 | 3.0 |
| 15 | $Zn_{40}Mg_{11}Ca_{31}Yb_{18}$ | 400 | 13.2 | 3.2 | 38 | $Ho_{55}Al_{25}Co_{20}$ | 651 | 12.6 | 3.0 |
| 16 | $La_{62}Al_{14}Cu_{20}Ag_4$ | 404 | 13.4 | 3.2 | 39 | $Zr_{65}Cu_{15}Ni_{10}Al_{10}$ | 652 | 12.8 | 3.1 |
| 17 | $Mg_{65}Cu_{25}Tb_{10}$ | 413 | 13.5 | 3.2 | 40 | $Zr_{55}Al_{10}Ni_5Cu_{30}$ | 682 | 12.8 | 3.1 |
| 18 | $Mg_{65}Cu_{25}Sm_{10}$ | 416 | 13.2 | 3.2 | 41 | $Zr_{55}Cu_{25}Ni_{10}Al_{10}$ | 685 | 13.3 | 3.2 |
| 19 | $Mg_{65}Cu_{25}Gd_{10}$ | 417 | 13.3 | 3.2 | 42 | $Cu_{46}Zr_{46}Al_7Gd_1$ | 700 | 12.9 | 3.1 |
| 20 | $Mg_{65}Cu_{25}Ho_{10}$ | 417 | 12.6 | 3.0 | 43 | $Cu_{46}Zr_{46}Al_8$ | 701 | 13.3 | 3.2 |
| 21 | $Mg_{65}Cu_{25}Y_{10}$ | 418 | 12.9 | 3.1 | 44 | $Zr_{50.7}Cu_{28}Ni_9Al_{12.3}$ | 719 | 12.9 | 3.1 |
| 22 | $La_{57.6}Al_{17.5}Cu_{12.4}Ni_{12.5}$ | 435 | 13.1 | 3.2 | 45 | $Zr_{44}Cu_{44}Al_6Ag_6$ | 722 | 13.1 | 3.2 |
| 23 | $La_{55}Al_{25}Ni_5Cu_{10}Co_5$ | 455 | 12.8 | 3.1 | | | | | |